%% file: main.tex
\documentclass[acmlarge]{acmart}

\def\etal{\emph{et al}.}
\def\vs{\emph{v.s.} }
\def\eg{\emph{e.g.}}
\def\ie{\emph{i.e.}}

\usepackage{xcolor,colortbl}
\usepackage{fancyhdr}

\definecolor{green}{HTML}{008015}

\definecolor{lightgray}{rgb}{0.9, 0.9, 0.9}

\definecolor{high}{HTML}{ecae00}
\definecolor{limited}{HTML}{a6a761}
\definecolor{minimal}{HTML}{8c7845}

\usepackage{array, multirow, makecell, booktabs}

\usepackage{tcolorbox}
\usepackage{graphicx}
\usepackage{enumitem}
\usepackage{wrapfig}
\usepackage{caption}
\tcbuselibrary{skins}
\usepackage{tikz}
\usepackage{adjustbox}
\usepackage{float}
\usepackage{fancyhdr}

\definecolor{highrisk}{RGB}{220, 53, 69}
\definecolor{limitedrisk}{RGB}{255, 193, 7}
\definecolor{minimalrisk}{RGB}{40, 167, 69}

\newtcolorbox{usecasecard}[2][]{
  colback=white,
  colframe=#2,
  fonttitle=\bfseries\large,
  title=#1,
  boxrule=1pt,
  arc=2mm,
  left=4pt,
  right=4pt,
  top=4pt,
  bottom=2pt,
  before skip=2pt,
  after skip=15pt,
  before upper={\setlength{\baselineskip}{1.2\baselineskip}\selectfont},
}

\definecolor{lightred}{RGB}{220, 100, 100}
\definecolor{amber}{RGB}{210, 140, 0}
\definecolor{pistachio}{RGB}{100, 160, 80}
\definecolor{brown}{RGB}{150, 90, 60}

\usepackage{soul}

\newenvironment{usecasecardlayout}[2]{
  \begin{usecasecard}[#1]{#2}
}{
  \end{usecasecard}
}

\newcommand{\cardsection}[1]{\noindent\textbf{#1}\par\smallskip}

\usepackage{todonotes}

\setcopyright{none}
\renewcommand\footnotetextcopyrightpermission[1]{}

\makeatletter
\AtBeginDocument{%
  \renewcommand{\ps@firstpagestyle}{%
    \let\@oddfoot\@empty
    \let\@evenfoot\@empty
    \def\@oddhead{}
    \def\@evenhead{}
  }%
  \renewcommand{\ps@standardpagestyle}{%
    \let\@oddfoot\@empty
    \let\@evenfoot\@empty
    \def\@oddhead{\hfill\thepage\hfill}
    \def\@evenhead{\hfill\thepage\hfill}
  }%
}
\makeatother

\makeatletter
\AtBeginDocument{%
  \def\@shortauthors{}%
  \def\@headfootfont{\small}%
  \fancyhf{}%
  \fancyfoot[C]{\thepage}%
}
\def\ACM@linecountL{}
\def\ACM@linecountR{}

\newcommand\blfootnote[1]{%
  \begingroup
  \renewcommand\thefootnote{}\footnote{#1}%
  \addtocounter{footnote}{-1}%
  \endgroup
}

\begin{document}

\title{Context Matters: Auditing Gender Bias in T2I Generation through Risk-Tiered Use-Case Profiles}

\author{Jose Luna}
\authornote{Equal contribution.}

\affiliation{%
  \institution{Singapore Management University}
  \country{Singapore}
}

\author{Yankun Wu}
\authornotemark[1]            

\affiliation{%
  \institution{The University of Osaka}
  \country{Japan}
}

\author{Xiaofei Xie}

\affiliation{%
  \institution{Singapore Management University}
  \country{Singapore}
}

\author{Noa Garcia}

\affiliation{%
  \institution{The University of Osaka}
  \country{Japan}
}

\renewcommand{\shortauthors}{Luna et al.}

\begin{abstract}
Text-to-image (T2I) generative models are increasingly used to produce content for education, media, and public-facing communication, and are starting to be integrated into higher-impact pipelines. Since generated images tend to reinforce stereotypes, producing representational erasure via ``default'' depictions and shaping perceptions of who belongs in certain roles, a growing body of work has proposed metrics to quantify gender bias in T2I outputs. Yet existing evaluations remain fragmented. Metrics are often reported without a shared view of what they measure, what assumptions they entail, or how their results should be interpreted under different deployment contexts. This limits the usefulness of gender bias measurement for both technical auditing and emerging governance discussions. We propose a risk-aligned auditing framework for gender bias in T2I models composed of three constituents that connects risk categories, evaluation metrics, and harms. First, we identify \emph{risk-tiered use-case profiles} aligned with the EU AI Act’s risk categories to motivate why auditing expectations may vary with deployment contexts and stakeholder exposure. Second, we construct a \emph{metric catalog} that consolidates gender-bias evaluation methods and organizes them in three measurement categories: gender prediction, embedding similarity, and downstream task. Third, we introduce a \emph{harm typology} that maps context-dependent harm categories (\eg, representational, quality-of-service) to specific risk-tired scenarios. Finally, we introduce THUMB cards (\underline{T}ext-to-image \underline{H}arms-informed \underline{U}se-case-aligned \underline{M}etrics of gender \underline{B}ias) that help formulate auditing systematically by the incorporation of context, scenario and bias manifestation, harm hypotheses, and audit strategy. Thus, by linking measurements and harms to specific risk-tiered deployment contexts, our framework supports a more reusable format for better auditing reporting.

\end{abstract}

\maketitle

\section{Introduction}
\blfootnote{\textcolor{black}{This work has been accepted for publication at the ACM FAccT 2026.}}

The rapid proliferation of text-to-image (T2I) generative models has fundamentally transformed creative production, content generation, and visual communication across industries. Systems such as DALL-E \cite{betker2023improving}, Stable Diffusion \cite{podell2023sdxl}, and Midjourney \cite{midjourney2025} have demonstrated remarkable capabilities in synthesizing photorealistic images from natural language prompts, enabling applications spanning entertainment, advertising, education, and beyond \cite{ko2023large,zhao2024enhancing,vartiainen2023using}. Yet this technological advancement has faced critical concerns \citep{bird2023typology,katirai2024situating} including, but not limited to, cultural bias \citep{elsharif2025cultural}, privacy \cite{carlini2023extracting}, and copyright \citep{somepalli2023diffusion}. As T2I generative models have been shown to systematically encode and amplify societal biases \cite{bianchi2023easily,luccioni2023stable}, their broad application may result in stereotypical representations of the world, reinforcing inequitable social norms. 

Across societal biases, gender is a particular salient and policy relevant for gender equality \citep{lutz2024ai,eu_ai_act_recital48_2024}. This salience is grounded not only in non-discrimination and accountability principles \cite{oecd2024principle}, but also in empirical evidence that T2I models exhibit gender defaulting and stereotyping patterns \citep{bianchi2023easily,naik2023social}. Additionally, the potential harms of gender-biased Artificial Intelligence (AI) have been illustrated in other applications such as in Natural Language Programming (NLP), recommender systems, and others  \citep{shrestha2022exploring,huang2025bias}. AI systems deployed in high-stakes settings such as facial recognition or recruitment have exhibited alarming gender biases \cite{buolamwini2018gender,njoto2022gender}.  
Likewise, UNESCO reveals evidence of regressive gender stereotypes \cite{UNESCO2024GenerativeAIStudy}. 
These examples underscore that if T2I models are integrated into content generation pipelines, gender inequalities may be reinforced at scale. Recognizing this risk, policymakers are pushing for greater accountability and transparency in AI. The European Union’s AI Act \cite{EU2024AIAct} establishes a risk-based regulation and presents a risk-tier categorization covering unacceptable, high, limited, and minimal risk. T2I models would be considered under general-purpose AI, where their gender bias risk could fall under either of the four categories depending on their deployment context. In light of arising AI regulations and governmental guidelines, there is both ethical and legal impetus to audit and mitigate biases in T2I before they manifest as greater societal harms.

However, auditing bias in T2I generation presents unique challenges, mainly due to the output modality and the lack of a standard bias definition itself. On one hand, unlike classification or decision-making systems that produce discrete labels, T2I models output images, which are  more complex to analyze due to their high dimensionality and unique combination of aesthetics, semantics, and compositions. Additionally, image generation lacks a standard definition of gender bias, often operationalized and limited to disparities in gender representations \cite{lee2023holistic, bakr2023hrs, cho2023dall} or misalignments between generated content and real-world distributions \cite{luccioni2023stable, bianchi2023easily}. Recent research has begun to examine these biases through various lenses. Most existing studies focus on analyzing the predicted gender in generated faces \cite{cho2023dall, li2024self}, or the distribution of representations in the latent space \cite{wang2023t2iat, ghosh2023person, wu2024stable}, while others evaluate disparities in downstream task performance when using the generated images as input \cite{garcia2023uncurated, ghosh2024don, ghosh2023person}. However, each approach has limitations. For instance, identifying the gender of a generated face typically relies on downstream classifiers or human judgment, which can reinforce preconceived norms of gender expression \cite{teo2023measuring}. Moreover, bias in generative models is not limited on the facial attributes and it can be manifested across the entire generated image, including objects, background, and image composition \cite{wu2024stable}. In such cases, evaluating only a portion of the image is insufficient to fully capture the scope of the bias presented in the output.

To make the complexities of gender-bias evaluation more accessible, our work seeks to provide a stepping stone towards auditing gender bias in T2I models. To do so, we provide an auditing framework operationalized by THUMB cards (\underline{T}ext-to-image \underline{H}arms-informed \underline{U}se-case-aligned \underline{M}etrics of gender \underline{B}ias)  
as a reusable audit-reporting format that i) aligns gender-bias assessment with regulatory risk framing, ii) consolidates concrete evaluation metrics, and iii) foregrounds context-dependent harms, bridging technical rigor and practical applicability.
Our contributions are summarized as follows:

\begin{enumerate}
    \item \textbf{Operationalizing risk-tiered auditing for T2I gender bias.} We instantiate EU AI Act-informed use-case profiles that translate risk tiers into concrete T2I deployment scenarios, specifying how gender bias is likely to manifest under each tier.
    \item \textbf{A structured metric catalog for gender bias in T2I.} We curate and categorize evaluation metrics into a unified catalog that improves comparability and supports informed metric selection by making explicit each method’s measurement target and intended use.
    \item \textbf{A context-grounded harm typology.} We derive a harm typology tailored to T2I gender bias and connect it to scenario-driven harm hypotheses, enabling auditors to interpret metric outputs in terms of plausible real-world harms rather than metric values alone.
    \item \textbf{THUMB cards for reusable audit reporting.} We introduce THUMB cards, a standardized reporting artifact that instantiates our framework into audit-ready units by combining deployment context and scenario-driven bias manifestations, explicit harm hypotheses, and a metric-grounded audit strategy, thereby supporting more systematic and reusable gender-bias audit documentation across risk tiers.
\end{enumerate}

In summary, this work connects gender bias evaluation for T2I generation and practical use case profiles aligned with the EU AI Act, providing an audit framework that operationalize these links through THUMB cards. Our contributions aid on the selection of appropriate metrics to assess and reduce potential harm eroding from different deployment contexts. Moreover, the proposed THUMB cards are designed as a practical, structured documentation artifact for key stakeholders, including developers, product teams, and internal or third-party auditors who must plan, document, and justify bias testing choices in a way that is traceable and context-proportionate. The cards provide a low-friction means of translating risk-tier expectations into auditable evidence by documenting what was tested, why particular metrics were selected, which harms were considered, and the deployment context in which the evaluation was conducted. To the best of our knowledge, this is among the first frameworks for auditing T2I gender bias that aligns with the EU AI Act’s risk categories and accounts for deployment context and harm consideration. By uniting diverse technical approaches under a common, risk-oriented structure, we hope to accelerate progress toward fair and inclusive generative models that align with societal values and emerging regulatory norms.

\section{Background \& Related Work}
This section reviews the related work needed to situate gender-bias auditing as both a technical and a governance problem. Because bias evaluation in text-to-image systems depends not only on how bias manifestations are identified, but also on how harms are interpreted and how evidence is documented for accountability, we draw on complementary literatures on AI bias evaluation, sociotechnical risk and harm, and audit-oriented governance. We therefore first review academic work on bias in AI and then examine relevant governance guidelines.

\subsection{Academic literature on AI bias evaluation, risk, and harm}

Recent work has increasingly examined biases in T2I outputs. \citet{sun2024smiling} audited DALL·E 2’s depictions of 153 occupations and found representational bias with gender distribution reflecting and amplifying real-world stereotypes. 
Moreover, \citet{ghosh2024don} revealed a large disconnect between users’ expectations and actual Stable Diffusion outputs, where even simple prompts can yield biased or stereotypical representations, particularly affecting marginalized groups. Beyond these patterns, research has uncovered novel representational harms, including exoticism and cultural misappropriation in depictions of non-Western contexts \cite{ghosh2024generative,liu2024scoft}. These studies call for more inclusive, culturally-aware T2I design underscoring the value of involving affected communities in bias evaluation. 

Furthermore, several studies have focused on how to investigate and mitigate AI risks, including bias, in ways aligned with emerging regulations. For instance, \citet{panigutti2025investigate} proposes a taxonomy of study designs for auditing algorithmic harms under the EU DSA’s risk management approach \citep{JRC2025AuditingAlgorithms}, while \citet{rao2025ai} analyzes model cards to catalog reported risks, underscoring the need for structured reporting that links technical bias metrics to real-world harms. At the same time, current AI ethics evaluation remains fragmented. Reviewing 800 metrics, \citet{rismani2025measuring} finds that most focus on a limited set of principles at isolated system levels, warning that evaluations often ``measure in pieces'' rather than capturing how harms emerge in context. In the T2I domain, \citet{bird2023typology} similarly develops a typology of risks, from data bias to model misuse, and highlights major governance and research gaps.
Additionally, scholars are examining how AI risk narratives shape governance priorities. \citet{oldenburg2025stories} analyze three competing ``risk imaginaries'' (the existential risk proponents, accelerationists, and critical AI scholars), showing each narrative frames AI risks differently and can shape policy agendas by sidelining alternatives. They also urge a pragmatic, harm-based approach to AI governance. Our work is firmly in line with this perspective. By focusing on measurable, present-day harms like gender bias and aligning our audit framework with the EU’s risk-based regulatory schema, we prioritize tangible fairness issues over speculative risks. We take notice of the lessons presented by prior work by not only providing appropriate metrics for measuring gender bias but also considering the contextual risk in our case studies. By categorizing use-case profiles per the EU AI Act risk levels, we aim to align bias audits with the gravity of potential harms. In short, prior work, from broad taxonomies of harms to detailed bias evaluations, collectively underscores the importance of transparent evaluation \cite{mitchell2019model}.

\begin{figure}
    \centering
    \hspace{20pt}
    \includegraphics[width=\linewidth]{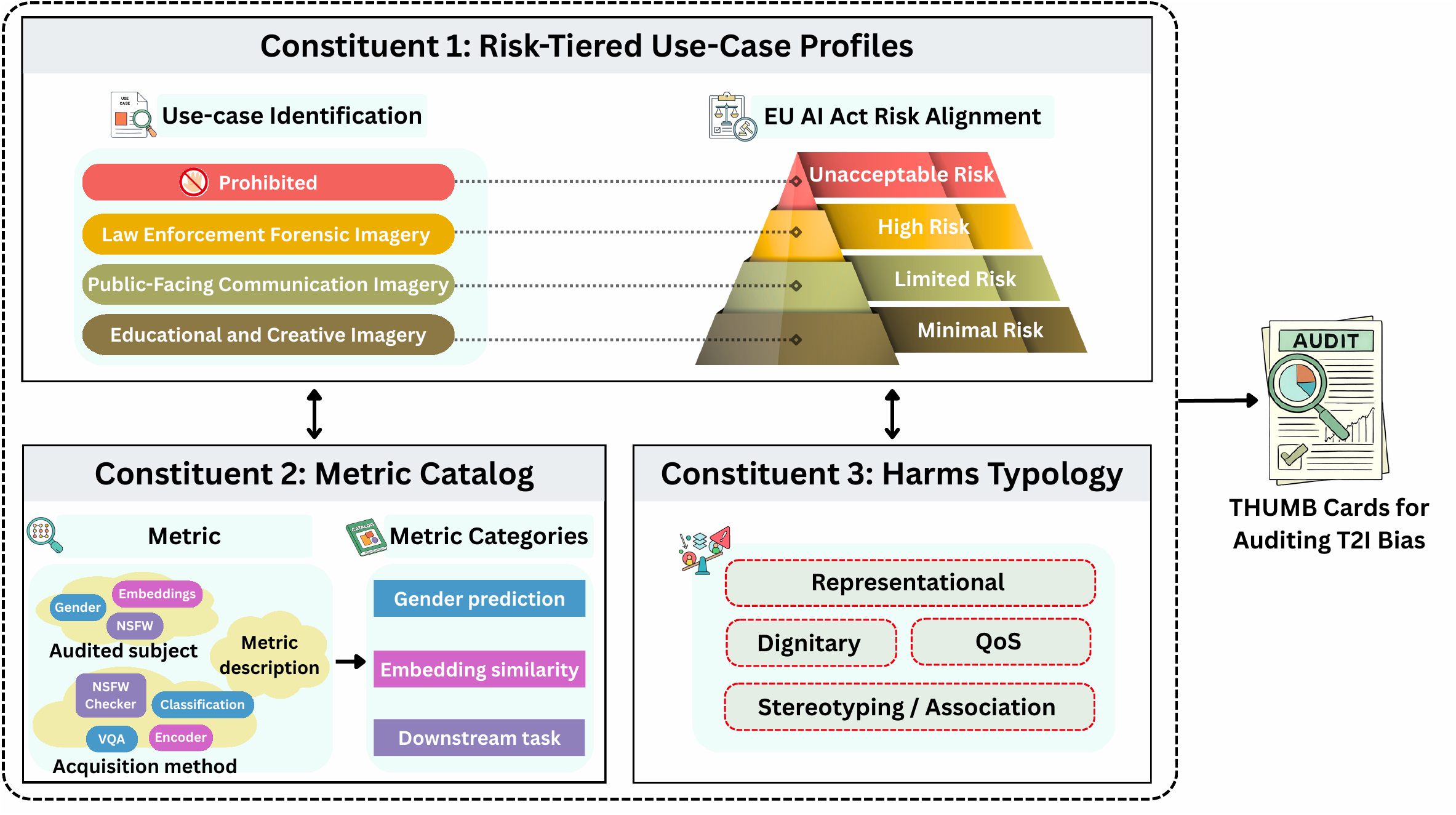}
    \caption{Overview of the T2I auditing framework, which is formed by three constituents and operationalized through the proposed THUMB Cards. ``QoS'' means Quality-of-Service.}
    \label{fig:methodology}
\end{figure}

\subsection{Governance-oriented guidance on AI bias auditing}

Growing concern about AI bias is also reflected in governmental guidelines, policy frameworks, and emerging standards. At the international level, the OECD AI Principles \cite{oecd_ai_principles_p6} call for human rights, fairness, and non-discrimination under \textit{Principle 1.2}, while UNESCO likewise urges the elimination of bias in AI systems to support inclusive and equitable AI \cite{UNESCO2021EthicsAI}. However, these documents remain high-level and do not provide concrete auditing or measurement methods.

Governmental efforts have also emerged, for example, the EU AI Act \cite{EU2024AIAct} adopts a risk-based approach to AI oversight, imposes obligations on high-risk systems, and offers broader governance coverage than principle-, rule-, or outcomes-based approaches \cite{luna2024navigating}. Under the Act, providers of high-risk systems must implement rigorous risk management and data governance measures to support fairness, while general-purpose AI providers are also expected to mitigate bias and safeguard fundamental rights. Outside the EU, non-binding guidelines that set expectations for bias testing have been issued. In the United States, the ``Blueprint for an AI Bill of Rights'' \cite{WhiteHouse2022AIBillOfRights} includes ``Algorithmic Discrimination Protections'', calling for the prevention and mitigation of discrimination, including gender bias. These principles are further reflected in OMB Memorandum M-24-10 \cite{OMB2024M2410}, which directs federal agencies to manage AI-related risks, including bias affecting public rights and safety.

Standards bodies have also issued practical frameworks. 
The NIST AI Risk Management Framework \cite{NIST2023AIRMF} identifies harmful bias as a core dimension of trustworthiness and calls for bias identification, measurement, and mitigation across the AI lifecycle. Similarly, NIST Special Publication 1270 \cite{NIST2022SP1270} examines sources of bias and provides a taxonomy for bias testing, evaluation, verification, and validation (TEVV), emphasizing that bias is a socio-technical, context-dependent issue requiring systematic audit processes.
Moreover, IEEE 7003 \cite{IEE7003} provides a framework for identifying, documenting, and mitigating algorithmic biases, encouraging practices such as bias impact assessment, stakeholder participation, and continuous auditing.
Furthermore, ISO/IEC TR 24027:2021 \cite{ISOIECTR24027_2021} catalogs bias types and sources and surveys techniques for their measurement and mitigation. Additionally, ISO/IEC 42001:2023 \cite{ISOIEC42001_2023} requires organizations to establish AI governance and risk assessment processes, including audits for fairness and non-discrimination. For certification, organizations must govern AI assets and assess technical and ethical risks, making gender bias a key concern to address.

In summary, both academic research and governance initiatives increasingly recognize that gender bias poses risks requiring measurement and management. Prior work has identified these risks and called for comprehensive responses, while regulators and standards recommend audits and risk analysis but provide only high-level guidance. We therefore align our framework with the EU AI Act’s risk categorization, given its maturity, broad policy influence, and binding obligations beginning to take effect from 2025 \cite{artificialintelligenceact2024timeline}. 
Moreover, our framework addresses the need for a specific auditing guideline by cataloging bias evaluation metrics and explaining how they can be applied to detect gender bias in T2I models. In doing so, it responds to policy demands for operational frameworks that support transparency, non-discrimination, and trustworthiness. We position our contribution as both novel and urgently needed to connect the dots between T2I fairness research and effective AI governance.

\section{Framework Construction}
\label{methodology section}
The construction of our framework follows three parts, each of them leading to a constituent: \emph{Constituent 1 — Risk-tiered use-case profiles}, \emph{Constituent 2 — Metric catalog}, and \emph{Constituent 3 — Harm typology}. Together, they form the proposed gender bias auditing framework for T2I generation. 
The framework composition is illustrated in Figure \ref{fig:methodology}, and these three constituents are later operationalized through the use of our proposed THUMB cards.
In addition, Table~\ref{tab:harm-metrics} highlights how each constituent could be utilize for representative examples per risk-tier. 
Please refer to \ref{part_one_appendix} for additional positioning of our work. We next describe each constituent in turn.

\input{table/harm-metrics}

\subsection{Constituent 1 — Risk-tiered use-case profiles}  
\label{sec:risks}
This constituent is the first building block of our framework. The goal here is to align with the EU AI act's risk-tiered framing \cite{eu_ai_act_2024}. To do so, we provide specific use-case profiles for each of the Act' risk-tiers. The use-case profiles are mapped by aligning with the act's description and provided use-case examples \citep{ai_act_summary2025}. Moreover, by following this classification, Constituent 1 illustrates an important premise of our paper: for T2I systems, the same model-level gender bias signal can have different governance significance depending on where and how the system is used. The EU AI Act classifies risk at the level of \emph{intended use and potential impact}, whereas gender bias may be seen as a \emph{failure mode} whose harms are mediated by audience reach, decision relevance, and institutional setting. Below the specific risk-tiers and use-case profiles:

{\small
\begin{tcolorbox}[
    colback=gray!8,
    colframe=gray!50,
    arc=3pt,
    boxrule=0.2pt,
    left=6pt, right=6pt, top=4pt, bottom=4pt
]
\begin{itemize}[leftmargin=*, itemsep=1pt]
    \item \textbf{Unacceptable Risk:} AI practices that pose clear threats to safety, rights, or livelihoods (\eg, AI-enabled real-time remote biometric identification systems). Systems in this category are banned \citep{eu_ai_act_2024_article5} and are therefore excluded.\\
    {\textcolor{red!60}{$\boldsymbol{\hookrightarrow}$}} \textsc{Use-case profile:} \textit{No Applicable}.

    \item \textbf{High Risk:} AI systems used in regulated products or sensitive domains (\eg, law enforcement) where failures can harm people, triggering the Act's strict requirements.\\
    {\textcolor{orange!70}{$\boldsymbol{\hookrightarrow}$}} \textsc{Use-case profile:} \textit{Law Enforcement Forensic Imagery}.

    \item \textbf{Limited Risk:} AI uses that are generally permitted but require transparency disclosures (\eg, users must be told they are interacting with AI or viewing AI-generated/manipulated content).\\
    {\textcolor{brown!70}{$\boldsymbol{\hookrightarrow}$}} \textsc{Use-case profile:} \textit{Public-Facing Communication Imagery (Media \& Health)}.

    \item \textbf{Minimal Risk:} All other AI uses that fall outside the previous categories.\\
    {\textcolor{green!60!black!60}{$\boldsymbol{\hookrightarrow}$}} \textsc{Use-case profile:} \textit{Educational and Creative Imagery}.
\end{itemize}
\end{tcolorbox}
}

At last, Constituent~1 is the connective tissue of the framework. It links Constituent~2 and Constituent~3 (described next) by turning them into a disciplined selection logic. Deployment context (from each use-case profile) determines which bias manifestations matter most, those manifestations motivate contextual \emph{harm hypotheses}, and harm hypotheses in turn facilitate which \emph{metric categories} are needed. Concretely, varying the application context while holding the gender bias dimension fixed, provides a principled way to explain why auditing expectations differ across risk-categories. For example, why high-risk settings warrant corroborative, multi-category metrics and tighter protocol documentation, whereas minimal-risk settings may justify lighter-weight checks and escalation triggers.

\subsection{Constituent 2 — Metric catalog}  
\label{sec:metrics}
\vspace{2mm}
The goal of this component is to provide a united view of (often fragmented) metrics to aid in the selection of appropriate evaluation methods. Thus, we construct a structured catalog of gender-bias evaluation methods for T2I models by consolidating metrics and evaluation protocols proposed in top-tier peer-reviewed venues in both the deep/machine learning and the fairness fields in the past 3 years.\footnote{Selected venues: NeurIPS, ICLR, CVPR, ECCV, ICCV, FAccT, AIES, ACL, EMNLP. See Table \ref{tab:supp_survey} for a detailed list of included papers.} The goal is not to claim an exhaustive census of all possible gender-bias measurements, but to provide a transparent and audit-oriented account of the methods that practitioners may encounter. 
This classification is intended as an audit aid rather than a normative claim about which metric category is the ``best''. The metric catalog is provided in Table~\ref{tab:survey}, where each row is a different \emph{evaluation method}, \ie, a concrete measurement procedure. This structured view can help researchers and auditors select appropriate metrics for auditing T2I models aligned with the EU AI Act risk tiers.

To construct the catalog, we first included studies that (i) evaluate images generated from T2I models, and (ii) operationalize gender-related bias or disparity through a measurable evaluation procedure. For each selected paper, we identify the procedure that quantifies bias as an evaluation method. 
When multiple evaluation methods are used within the same paper, each procedure is recorded as a separate entry in the catalog. Each evaluation method is then systematically coded according to a set of descriptive attributes that reflect how bias manifests and capture the components of the evaluation process. These attributes characterize what part of the generated image is analyzed, what subject of bias is audited, and what technique is used to measure bias. Using this coding scheme allows various evaluation protocols reported across different papers to be represented within a common analytical structure.

Based on the coded evaluation procedures, we then derived three \emph{metric categories} that reflect different aspects of bias evaluation: \emph{gender prediction}, \emph{embedding similarity}, and \emph{downstream task}. These categories reflect different ways in which bias is assessed — either through predicted attributes, similarities in latent spaces, or the effect of generated images when processed by downstream models. 
Importantly, this grouping reflects how prior studies have demonstrated gender bias evaluation in T2I generation. The catalog should not be read as a normative ranking of metrics or as an exhaustive categorization, but rather as a functional organization of evaluation strategies to understand what type of evidence each method produces and when it may be applicable.

\subsubsection{Metric catalog attributes} 
\vspace{2mm}
Table~\ref{tab:survey} summarizes evaluations of gender bias in T2I generation, and assigns each method the following attributes: 
\vspace{4mm}

\begin{itemize}
    \item \textbf{Image type.} The type of generated images based on their composition. Gender bias evaluation generally focuses on two types of images: 
    \begin{itemize}
    \item \textit{Face.} Face-centric generated images with little room to depict a background or other objects \cite{teo2023measuring, shi2025dissecting}. 
    \item \textit{Scene.} Images depicting a full scene including people, objects, backgrounds, and/or other elements \cite{d2024openbias, sathe2024unified}.
    \end{itemize}
    \item \textbf{Metric category.} Each \emph{metric category} targets a specific aspect of gender bias evaluation. We group metrics into three categories based on the means to conduct the evaluation:
    \begin{itemize}
        \item \textit{Gender prediction.} Metrics that evaluate gender bias by first predicting the gender in the generated image, and then quantifying statistical disparities based on the predicted gender.
        \item \textit{Embedding similarity}. Metrics that evaluate gender bias based on the image embeddings or the intermediate embeddings during the generation, rather than a predicted gender label. This avoids direct reliance on discrete labels and captures implicit stereotypes that may not be revealed by gender prediction metrics.
        \item \textit{Downstream task.} Metrics that focus on the performance of downstream tasks/applications when generated images are used. These methods measure gender-related disparities through downstream performance or application-facing outcomes (\eg, differential NSFW rates \cite{garcia2023uncurated, ghosh2023person}), linking bias to functional impacts.
    \end{itemize}
    \item \textbf{Audited subject.} The subject that is evaluated to measure gender bias under a given metric. For example, gender prediction metrics audit gender labels, while embedding similarity metrics audit image embeddings. In downstream task metrics, the audited subject depends on the specific focus. For instance, if the metric aims to evaluate gender disparities in NSFW content, the NSFW label would serve as the audited subject.
    \item \textbf{Acquisition method.} The technical process used to obtain the audited subject. For instance, in gender prediction metrics, the gender label, which is the audited subject, may be predicted using a classifier \cite{karkkainen2021fairface} or extracted from generated textual descriptions from an image captioning model \cite{li2023blip}. 
    \item \textbf{Metric description.} Brief text description of the evaluation method. 
\end{itemize}

\vspace{2mm}

\input{table/survey}

\subsubsection{How these metrics inform EU AI Act risk-based auditing} Under the EU AI Act, systems with minimal or limited risk may not require extensive bias audits, but basic distribution metrics can still be used to flag obvious gender skew. For example, simple metrics that measure the proportion of \textit{female vs male} can reveal if a model heavily favors one gender in its outputs, a concern even for consumer-facing generative models from an ethical standpoint. In higher risk applications (\eg, imagery used in law enforcement), regulators may expect more comprehensive bias evaluations. Here, performance parity metrics \cite{lee2023holistic, bakr2023hrs} are crucial – ensuring the model performs equally well for all genders helps prevent indirect discrimination (a key requirement for high-risk systems). Similarly, implicit association tests \cite{wang2023t2iat} reveal subtler biases (\eg, consistently linking women to family roles and men to career roles). Such biases might not be evident from overall proportions but could still produce harmful stereotyping, which is vital to uncover in high-risk scenarios. Metrics like T2IAT \cite{wang2023t2iat} or those comparing outputs to real-world data are thus better suited for high-risk system auditing.
Additionally, safety-related bias metrics (\eg, NSFW flag disparities \cite{garcia2023uncurated, ghosh2024don}) are important for content generation models; if a model’s safety filter flags one gender’s images more often, this could raise fairness and dignity concerns. For an unacceptable risk use-case (which are prohibited uses of AI under the Act, such as social scoring or overtly harmful AI), any significant bias revealed by these metrics would reinforce the system’s unsuitability. In summary, auditors should employ a combination of these metrics proportional to the system’s risk level.

\subsection{Constituent 3 — Harm typology} 
\label{sec:harms}
\vspace{3mm}
This constituent helps complimenting the auditing framework by mapping to the \emph{risk-tiered use-case profiles}, and consequently to the \emph{metric catalog}. For a given use-case profile, we specify which gender-bias-related harms are plausible to consider, how such harms would manifest in generated imagery, and which metric categories are more appropriate. Our aim is not to provide an exhaustive social theory of gender harm, but to derive an audit-oriented typology that is sufficiently grounded in fairness scholarship while remaining actionable for T2I evaluation. To do so, we drew candidate harm categories from two bodies of work. First, algorithmic fairness scholarship that distinguishes harms in representation \citep{barocas2023fairness}; and second, sociotechnical harm scholarship that emphasizes that AI harms may also arise through dignity violations, performance disparities, and broader downstream social effects \citep{shelby2023sociotechnical}. We then compared these candidate categories against the concrete gender-bias manifestations documented in T2I studies, including representational skew and erasure in occupational imagery \citep{sun2024smiling,bianchi2023easily,wu2024stable,wang2024new}, stereotypical role associations and gendered portrayals \citep{bird2023typology,wu2024stable}, demeaning or objectifying depictions \citep{ghosh2024don,asim2026through,castleman2025adultification}, unequal task performance or prompt fidelity across genders \citep{lee2023holistic,ghosh2024don}, and downstream effects on viewers’ beliefs and implicit bias \citep{guilbeault2024online,sim2025biased}.\\

We retained a harm category only when it satisfied three conditions. First, it had to be conceptually established in academic fairness or sociotechnical-harm literature, so that the category was not ad hoc. Second, it had to be empirically observable in existing T2I gender-bias research, so that it corresponds to documented bias manifestations rather than speculative concerns. Third, it had to be audit-useful, meaning that it helps differentiate what kind of evidence should be collected for a given deployment context. This procedure yielded five categories: representational harms, stereotyping harms, dignitary harms, quality-of-service harms, and downstream allocative impact. Together, these categories span five analytically distinct questions relevant to auditing— \textit{who is shown}, \textit{how they are portrayed}, \textit{whether the portrayal is degrading}, \textit{whether the model performs reliably across genders}, and \textit{whether the outputs can plausibly influence consequential judgments or opportunity-relevant beliefs}. Below, a description of each harm is given:\\

\begin{itemize}
    \item \textbf{Representational harms.} Harms that arise when T2I outputs systematically skew \emph{who is depicted} across prompts, shaping the visibility of genders in ways that can normalize exclusion or erasure \citep{barocas2023fairness}. As example, systematic over-/under-representation and potential \emph{erasure} across prompts (\eg, gender-neutral prompts disproportionately ignoring one gender).
    \item \textbf{Stereotyping harms.} Harms that arise when outputs reproduce or legitimize gender stereotypes by coupling gender with particular roles, traits, or visual codes \cite{bird2023typology}. They can be seen as patterned gender–attribute dependencies in \emph{how gender is depicted} in, for example, gendered occupation.
    \item \textbf{Dignitary harms.} Affronts to dignity or self-respect caused by demeaning, derogatory, or objectifying imagery \citep{asim2026through, castleman2025adultification}. For example, derogatory or disproportionately sexualized depictions that threaten dignity and social standing in context. 
    \item \textbf{Quality-of-service (QoS) harms.} Group-based performance disparities where the system disproportionately under-performs for certain social groups \cite{shelby2023sociotechnical}. For example, unequal performance or utility across genders, such as differences in prompt adherence or alignment in outcomes.
    \item \textbf{Downstream allocative impact.} Context-dependent pathways whereby biased imagery can influence opportunity-relevant decisions, institutional actions, humans’ implicit gender bias, or viewers’ beliefs about who ``belongs'' in particular roles. We treat this category as \emph{conditional} on deployment, rather than as an inherent property of all T2I uses. For example, nowadays large volumes of generated imagery circulate online, and such online images can amplify gender bias by shaping who is exposed to these depictions \citep{guilbeault2024online}. 
\end{itemize}

These categories are intended as an audit scaffold. They reflect established fairness harm framings and sociotechnical accounts of harm \citep{shelby2023sociotechnical}, they align with bias signals that can be measured using existing T2I evaluation methods (Table~\ref{tab:survey}), and they can be instantiated as context-dependent hypotheses tied to concrete use-case profiles. Moreover, they are broad enough to generalized across deployment contexts, but specific enough to motivate different evidentiary strategies. For example, representational harms often call for distributional or gender-prediction metrics, stereotyping harms may require association-sensitive measures, dignitary and QoS harms often require downstream-task evidence, and downstream allocative impact typically requires a bundled interpretation tied to deployment context. Importantly, we do not infer real-world harm solely from metric deviations. Instead, we treat harms as \emph{hypotheses}. For each use-case profile, we specify the relevant harm hypotheses, describe the expected bias manifestations, and map each harm to one or more metric categories under each \emph{audit strategy}. We note that harms are not mutually exclusive, meaning, under real-world deployments, depictions may contain more than one harm type in a single image.\\

\vspace{9mm}
Section 3 has detailed the three constituents that form our framework, which is operationalized through the proposed THUMB Card presented in the next section. We emphasize that our work does not extend or reinterpret limitations of the EU AI Act; rather, it operationalizes the Act’s risk-tier logic to support context-proportionate auditing and documentation. In contrast to other AI-audit-related frameworks \citep{lee2023holistic, bakr2023hrs,ISOIEC24027}, which often overlook deployment context, we emphasize that context determines not only the suitability of particular evaluations but also, for organizations seeking to comply with the EU AI Act, how the risk level of a T2I-enabled application should shape the auditing approach. \\

\section{Framework operationalization: THUMB Cards}\label{use-cases section}
\vspace{2mm}
Building on our constituents, this section operationalizes the framework through THUMB cards. These are structured, reusable audit templates that translate risk-tier expectations into auditable evidence. The framework provides the rationale for audit design, while the cards record its application in a traceable form. This matters because auditing requires not only measuring bias, but also documenting why specific tests were chosen, what harms they are intended to probe or mitigate, and how findings should be interpreted given a specific context.\\

Each card specifies: (i) the deployment \emph{context}, (ii) a \emph{scenario and bias manifestation} in generated imagery (\eg, depiction defaults, stereotyped framing, or unequal prompt adherence), (iii) an illustrative \textit{harms example}, (iv) a set of \emph{harm hypotheses} conditioned on that context, and (v) an \emph{audit strategy} describing the recommended evaluation metrics. The cards are modular and adaptable as auditors may instantiate one profile, refine it to fit organizational needs, or introduce additional profiles for other domains while preserving the same traceability structure. We illustrate this operationalization through three \emph{use-case profiles} of gender bias in T2I generation: 1) high-risk use-case profile for law enforcement forensic imagery (Figure \ref{fig:high-risk-card}), 2) limited-risk use-case profile for public facing communication imagery in media and health communication (Figure \ref{fig:limited-risk-card}), and 3) minimal-risk use-case profile for educational and creative imagery (Figure \ref{fig:minimal-risk-card}). These profiles were selected in line with the EU AI Act's indications of which applications correspond to each risk-level. For instance, law enforcement is a direct example given in Annex III use cases \cite{ai_act_summary2025}. Using the cards rather than only narrative description is important because existing governance and reporting artifacts serve different purposes. High-level guidance and bias frameworks identify relevant principles, risks, or categories, but do not by themselves provide a compact unit for documenting a concrete bias audit. Likewise, broader disclosure artifacts such as model cards \citep{mitchell2019model, katirai2024situating} or risk catalogs \citep{rao2025ai} support system-level transparency, but are not designed to tie together a specific deployment context, scenario-level bias manifestation, harm hypothesis, and metric-grounded audit strategy in a single traceable artifact. THUMB cards fill this middle layer — they are narrower than model-wide disclosures and more operational than abstract governance guidance.\\

This card-based instantiation provides a standardized structure for documentation and reporting of audit findings. Developers and product teams can use it during design and pre-deployment testing to scope prompt sets, select proportionate evaluations, and document findings. Responsible AI, governance, and compliance teams can use it to assess whether the evidence is commensurate with deployment risk and associated harms, while internal or third-party auditors can use it as a compact traceability artifact for what was tested and under which assumptions. For small and medium-sized enterprises (SMEs) in particular, the cards offer a low-friction way to structure internal bias testing and retain documentation that can later support external review. Although we instantiate three profiles aligned with high-, limited-, and minimal-risk tiers, the structure is generalizable to other deployment contexts, bias dimensions, and harm hypotheses. Overall, T2I systems should be audited in relation to specific contexts, tasks, and risks rather than through a one-size-fits-all evaluation.

\input{table/card_high_risk} 
\input{table/card_limited_risk}
\input{table/card_mininal_risk}

\section{Conclusions \& Limitations}
Gender bias in T2I generation is persistent and context-sensitive. Similar bias manifestations can imply markedly different levels of concern depending on where and how a model is deployed. In this paper, we present a risk-tiered, harm-centric auditing framework for T2I generative models that bridges metric-driven evaluation and governance-oriented oversight, enabling more comprehensive and context-sensitive audits of gender bias.

Our framework constructs three domain-specific \emph{use-case profiles} of gender bias and aligns them with the EU AI Act’s risk categorization. Using this risk-oriented lens, we introduce a \emph{metric catalog} that consolidates and organizes gender-bias evaluation methods. 
We further present a \emph{harm typology} tailored to T2I generative content, covering \textit{representational}, \textit{stereotyping}, \textit{dignitary}, \textit{quality-of-service}, and \textit{downstream allocative impact} harms. This typology provides a shared vocabulary for articulating how and whom biased image generation may harm. Finally, to operationalize our framework, we introduced THUMB cards which provides a standardize structure for auditing and reporting. Taken together, these contributions support the growth of AI auditing by aligning technical bias assessments with ethical and regulatory imperatives. Our work demonstrates how fairness metrics – often criticized for being disconnected from practice – can be embedded within a risk-aware audit workflow. Under our framework, evaluation results are not merely seen as performance scores; they constitute evidence within an accountability narrative that can be interpreted by diverse stakeholders, from model builders to compliance officers. In addition, the proposed framework is intended to help developers, product teams, and auditors document and justify bias evaluation decisions, support internal assurance, and provide a basis for regulatory alignment and future compliance. By tightly coupling deployment context, evaluation metrics, and harm association, the THUMB cards render audit findings more interpretable, actionable, and policy-relevant than one-dimensional bias reporting. More broadly, our approach generalizes beyond gender bias in T2I and can serve as a blueprint for Generative AI auditing practice. As regulators and standards bodies in Europe and beyond formulate transparency and fairness requirements, the ideas presented here can inform audit guidance and best practices for responsible and accountable Generative AI deployment.

\textbf{Limitations.} While our framework offers a novel and governance-aligned auditing scaffold, it is primarily synthesis-driven and has not yet been validated as an end-to-end operational auditing procedure. Future work should therefore evaluate the framework in practice through user studies with developers, auditors, and policymakers.
We focus our instantiation on gender bias, extending the approach to additional protected attributes, intersectional harms, and other generative modalities is a natural next step, and the framework’s modular design is intended to support such expansion.

\begin{acks}
This research is supported by the National Research Foundation, Singapore, the Cyber Security Agency under its National Cybersecurity R\&D Programme (NCRP25-P04-TAICeN), and the JSPS KAKENHI No. 22K12091 and No. 23H00497. Any opinions, findings and conclusions or recommendations expressed in this material are those of the author(s) and do not reflect the views of the National Research Foundation, Singapore, the Cyber Security Agency of Singapore, or the JSPS KAKENHI.
\end{acks}

\section*{Endmatter}
No generative AI tools were used for the research design, metric collections, analysis, or creation of original content in this paper. Generative AI assistance, where used, was limited to post-writing proofreading (grammar and syntax). All contributions (including our research methodology, literature analysis, metric catalog, use-case profiles, harm typology, and the drafting of the core content) were completed by the authors.

\bibliographystyle{ACM-Reference-Format}
\bibliography{sample-base}


\input{appendix}

\end{document}

%% file: table/harm-metrics.tex
\begin{table}[t]
\centering
\setlength{\tabcolsep}{5pt}
\scalebox{0.8}{\begin{tabular}{p{7.5cm}p{2.5cm}p{2.7cm}p{5cm}}
\toprule
\textbf{Scenario examples} & \textbf{Harm typology} & \textbf{Metric category} & \textbf{Auditing strategy}
\\

\midrule
\rowcolor{high}
\multicolumn{4}{l}{\textcolor{white}{High-Risk Use-Case Profile: Law Enforcement Forensic Imagery)}} \\

When gender is specified, outputs contradict or distort gender presentation; gender-neutral prompts default to stereotyped depictions & Representational & Pred, Emb & Pred: proportion disparities \cite{luccioni2023stable, teo2023measuring, lee2023holistic, naik2023social, kang2025fairgen, kim2025rethinking}, deviation from target distributions \cite{kang2025fairgen, shi2025dissecting, jung2025multi}; Emb: embedding similarities \cite{wu2024stable, ghosh2023person} \\
Gendered portrayals that are humiliating, sexualized, or otherwise degrading in sensitive investigative contexts & Dignitary & Task & NSFW detector \cite{garcia2023uncurated, ghosh2023person}, Expectation mismatch \cite{ghosh2024don}\\
Unequal failure modes: prompt adherence or relevant attributes are less reliable for one gender, creating systematic distortions & Quality-of-service & Pred, Task & Pred: gender accuracy \cite{wan2024factuality, bakr2023hrs}; Task: prompt-image alignment \cite{lee2023holistic}, expectation mismatch \cite{ghosh2024don}, hidden attributes \cite{wu2024stable} \\
Images used to support investigative narratives may amplify gendered assumptions (suspect/victim portrayals skewed by stereotypes), potentially influencing decisions & Downstream allocative impact & Bundle (Pred/Emb/Task) & Pred: distribution deviation \cite{bakr2023hrs, cho2023dall, zhou2024association, li2025t2isafety}; Emb: association tests \cite{wang2023t2iat}; Task: impact on humans' bias \cite{sim2025biased} \\

\midrule
\rowcolor{limited}
\multicolumn{4}{l}{\textcolor{white}{Limited-Risk Use-Case Profile: Public-Facing Communication Imagery (Media \& Health)}} \\

Gendered stereotypes in depictions relevant to health/media roles (\eg, caregiving vs authority roles) & Stereotyping & Pred, Emb & Pred: distribution deviations \cite{bakr2023hrs, cho2023dall, zhou2024association, li2025t2isafety}; Emb: association tests \cite{wang2023t2iat} \\
Disproportionate sexualization or demeaning depiction for one gender under comparable prompts; higher likelihood of NSFW-leaning outcomes & Dignitary & Task & Group disparity in NSFW/sexualized outcomes \cite{garcia2023uncurated, ghosh2023person} \\
Unequal factual adherence to explicitly stated gender (\eg, prompt specifies \texttt{woman doctor} but output female \textit{nurse}) & Quality-of-service & Pred & Proportion disparities \cite{luccioni2023stable, teo2023measuring, lee2023holistic, naik2023social, kang2025fairgen, kim2025rethinking}, Factuality \cite{wan2024factuality} \\
Biased imagery in public communication may systematically shape perceived “who belongs” in roles, with indirect effects on opportunities (\eg, campaigns, outreach) & Downstream allocative impact & Bundle (Pred/Emb/Task) & Pred: distribution deviation \cite{bakr2023hrs, cho2023dall, zhou2024association, li2025t2isafety}; Emb: association tests \cite{wang2023t2iat}; Task: impact on humans' bias \cite{sim2025biased}\\

\midrule
\rowcolor{minimal}
\multicolumn{4}{l}{\textcolor{white}{Minimal-Risk Use-Case Profile: Educational and Creative Imagery)}} \\

Gender-neutral prompts yield skewed gender proportions; limited diversity in gender presentation & Representational & Pred, Emb & Pred: distribution deviations \cite{bakr2023hrs, cho2023dall, zhou2024association, li2025t2isafety}, real-world vs generated proportions \cite{luccioni2023stable, bianchi2023easily, naik2023social, dehdashtian2025oasis, kang2025fairgen}; Emb: biased representation in the entire image \cite{wu2024stable} \\
Gender is implicitly associated with professions (\eg, \textit{nurse}→female; \textit{engineer}→male), even when prompts are neutral & Stereotyping & Pred & Gender proportion \cite{bianchi2023easily, shen2024finetuning, li2024self, fu2025fairimagen, sreelatha2025respodiff, jung2025multi, kang2025fairgen, park2025fair, wan2025male} \\
In gaming industry female characters may be represented with less clothing than male characters & Dignitary  & Task & NSFW detector \cite{garcia2023uncurated, ghosh2023person} \\
Unequal prompt adherence or output quality across genders (\eg, \texttt{a female teacher, no smile} yields a smiling woman) & Quality-of-service & Task & Prompt-image alignment \cite{lee2023holistic}, Expectation mismatch \cite{ghosh2024don}, Factuality \cite{wan2024factuality} \\

\bottomrule
\end{tabular}}
\caption{For each highlighted risk-tiered use-case profile (Section \ref{sec:risks}), we provide representative scenarios, state the primary harm typology (Section \ref{sec:harms}), and map it to the relevant metric category and auditing metrics in our metric catalog (Section \ref{sec:metrics}). ``Pred'', ``Emb'', and ``Task'' denote three metrics categories: \textit{gender prediction}, \textit{embedding similarity}, and \textit{downstream task}, respectively.}
 
\label{tab:harm-metrics}
\vspace{-10mm}
\end{table}

%% file: table/survey.tex
 \begin{table*}[!h] 
\setlength{\tabcolsep}{3pt}
\scalebox{0.74}{\begin{tabular}{@{}p{3cm}p{1.1cm}p{2.7cm}p{1.6cm}p{3.1cm}p{9.1cm}@{}}
\toprule
\raisebox{-.9\height}{\textbf{Reference}} & \textbf{Image type} & \raisebox{-.9\height}{\textbf{Metric category}} & \textbf{Audited subject} & \raisebox{-.9\height}{\textbf{Acquisition method}} & \raisebox{-.9\height}{\textbf{Metric description}} \\
\midrule
Luccioni \etal~\cite{luccioni2023stable} & face & gender prediction & gender & image captioning \cite{offert2022sign}, VQA \cite{li2022blip} & generated \vs real-world on gender proportions \\
Teo \etal~\cite{teo2023measuring} & face & gender prediction & gender & classification \cite{teo2023measuring} & female \vs male on proportions \\
Lee \etal~\cite{lee2023holistic} & scene & downstream task & embeddings & encoder \cite{hessel2021clipscore}, human annotation & female \vs male on prompt-image alignment \\
Lee \etal~\cite{lee2023holistic} & face & gender prediction & gender & classification \cite{radford2021learning} & female \vs male on proportions \\
Garcia \etal~\cite{garcia2023uncurated} & scene & downstream task & NSFW & NSFW detector \cite{safetychecker} & female \vs male on percentage of image marking as NSFW \\
Bakr \etal~\cite{bakr2023hrs} & scene & gender prediction & gender & classification \cite{deng2019arcface, deng2020retinaface, rothe2015dex} & mean absolute deviation (MAD) of gender distribution \\
Bakr \etal~\cite{bakr2023hrs} & scene & downstream task & caption & image captioning \cite{li2023blip} & gender disparities of prompt-caption alignment \\
Cho \etal~\cite{cho2023dall} & face & gender prediction & gender & image captioning \cite{li2023blip} & MAD of gender distribution \\
Bianchi \etal~\cite{bianchi2023easily} & face & gender prediction & gender & classification \cite{radford2021learning} & generated \vs real-world on gender proportions \\
Naik \etal~\cite{naik2023social} & face & gender prediction & gender & human annotation & generated \vs real-world on gender proportions \\
Wang \etal~\cite{wang2023t2iat} & scene & embedding similarity & embeddings & encoder \cite{radford2021learning} & WEAT \cite{caliskan2017semantics}-based association score on two concepts (\eg, science \vs art) and genders \\
Ghosh \etal~\cite{ghosh2023person} & face & embedding similarity & embeddings & encoder \cite{radford2021learning} & similarity between neutral and gendered images \\
Ghosh \etal~\cite{ghosh2023person} & face & downstream task & NSFW & NSFW detector \cite{nsfw} & female \vs male on probability of classified as \textit{sexy} and \textit{neutral} \\
Zhou \etal~\cite{zhou2024association} & face & gender prediction & gender & Encoder~\cite{zhou2024association} & total variation distance between gender distributions \\
Shen \etal~\cite{shen2024finetuning} & face & gender prediction & gender & classification & female \vs male on proportions \\
Li \etal~\cite{li2024self} & face & gender prediction & gender & classification \cite{radford2021learning} & largest relative deviation of genders from the uniform distribution \\
D{'}Inc{\`a} \etal~\cite{d2024openbias} & scene & gender prediction & gender & VQA \cite{liu2024improved, liu2023visual} & entropy of two gender probability distributions\\
Chinchure \etal~\cite{chinchure2023tibet} & face & gender prediction & gender & VQA \cite{chen2023minigpt} & MAD between the images from original and counterfactual prompts \\
Wu \etal~\cite{wu2024stable} & scene & embedding similarity & embeddings & T2I model \cite{rombach2022high}, encoder \cite{he2016deep, radford2021learning, caron2021emerging, somepalli2023diffusion} & similarity between neutral and genders on intermediate embeddings during generation and image embeddings\\
Wu \etal~\cite{wu2024stable} & scene & downstream task & objects & T2I model \cite{rombach2022high}, visual grounding \cite{ren2024grounded} & gender disparities on object frequency in the images and in the proposed prompt-image dependency groups \\
Ghosh \etal~\cite{ghosh2024don} & face & downstream task & appearance & human annotation & similarity between generated gender and users' expectation \\
Wan \etal~\cite{wan2024factuality} & face & gender prediction & gender & classification \cite{karkkainen2021fairface} & accuracy of generated gender \\
Sathe \etal~\cite{sathe2024unified} & Scene & gender prediction & gender & image captioning \cite{li2023blip} & difference on generated gender frequency \\
Sim \etal~\cite{sim2025biased}  & face & gender prediction & gender & classification & female \vs male on proportions \\
Sim \etal~\cite{sim2025biased} & face & downstream task & - & - & implicit association task on participants who with \vs without exposure to generated images \\
Fu \etal~\cite{fu2025fairimagen} & face & gender prediction & gender & classification \cite{serengil2024benchmark} & female \vs male on distribution of frequency in each generated image \\
Sreelatha \etal~\cite{sreelatha2025respodiff} & face & gender prediction & gender & classification \cite{radford2021learning} & largest relative deviation of genders from the uniform distribution \\
Dehdashtian \etal~\cite{dehdashtian2025oasis} & face & gender prediction & gender & classification \cite{radford2021learning} & generated \vs real-world on gender proportions \\
Li \etal~\cite{li2025t2isafety} & scene & gender prediction & gender & VQA \cite{li2025t2isafety} & normalized KL divergence between the distributions of two groups \\
Shi \etal~\cite{shi2025dissecting} & face & gender prediction & gender & classification \cite{karkkainen2021fairface} & gender distribution \vs uniform distribution \\
Kim \etal~\cite{kim2025rethinking} & face & gender prediction & gender & classification \cite{radford2021learning} & female \vs male on proportion \\
Jung \etal~\cite{jung2025multi} & face & gender prediction & gender & classification & gender distribution \vs reference distribution \\
Jiang \etal~\cite{kang2025fairgen} & face & gender prediction & gender & classification & gender distribution \vs uniform distribution \\
Park \etal~\cite{park2025fair} & face & gender prediction & gender & classification \cite{radford2021learning} & gender distribution \vs uniform distribution \\
Friedrich \etal~\cite{friedrich2025multilingual} & face & gender prediction & gender & classification \cite{karkkainen2021fairface} & MAD of gender distribution \\
Wan \etal~\cite{wan2025male} & face & gender prediction & gender & human annotation & alignment on generated gender and steretyped gender \\
Kang \etal~\cite{kang2025fairgen} & scene & gender prediction & gender & VQA \cite{instructblipv2} & absolute difference between proportions of genearated gender and target ones \\
Saeed \etal~\cite{saeed2025beyond} & scene & gender prediction & gender & VQA \cite{li2023blip, liu2023visual, Qwen-VL} & female \vs male on proportions \\
\bottomrule
\end{tabular}}
\caption{Catalog of gender bias evaluation studies in text-to-image generation. According to the evaluated target and the auditing methods, we categorize these evaluation metrics into three categories: \textit{gender prediction}, \textit{embedding similarity}, and \textit{downstream task}. We included metrics and evaluation protocols published in top-tier peer-reviewed venues in both deep/machine learning and fairness field (\eg, CVPR, NeurIPS, FAccT, etc) from 2023 to 2025.}
\vspace{-15pt}
\label{tab:survey}
\end{table*}

%% file: table/card_high_risk.tex
\begin{figure}[H]
\resizebox{\textwidth}{\textheight}{
\begin{usecasecardlayout}{\textbf{High-Risk Use-Case Profile: Law Enforcement Forensic Imagery}}{black}
\label{card:high-risk}

\cardsection{Context} In security and law-enforcement settings, T2I models may be used to generate suspect depictions or forensic sketches from textual descriptions (\eg, witness accounts). The generated imagery may plausibly influence judgments, investigative narratives, or institutional actions, even if outputs are not determinative on their own. 

\vspace{5pt}
\begin{minipage}[t]{0.67\textwidth}
\setlength{\baselineskip}{1.2\baselineskip}\selectfont
\cardsection{Scenario and bias manifestation}
Investigators may prompt a T2I model to produce a portrait-like depiction from a description such as ``Provide the sketch of a person robbed me'' (see examples at right).
Gender bias can manifest as a \emph{defaulting effect}, where gender-neutral or underspecified prompts systematically yield male-presenting suspects due to learned associations between crime-related contexts and men \citep{garside2024accuracy,neil2024trans}. This constitutes a form of representational skew.
Bias may also appear through \emph{stereotype injection} as prompts associated with certain crimes or attributes (\eg, ``convicted'') when combined to gender, may  produce diverse outputs (see the example) where arresting postures are more aggressive towards men; and more revealing attire and ``submissive'' posture for women. 
These outputs risk reinforcing harmful priors (\eg, perpetrators are male by default) and may mislead investigations. Real-world examples include ``Forensic Sketch AIrtist'' that was met with alarms as experts warned it could worsen existing racial and gender biases in witness descriptions \citep{vice2023aipolicesketches,fortunato2022forensic}.
\end{minipage}
\hfill
\noindent\begin{minipage}[t]{0.3\textwidth}
\cardsection{Harms example}

\vspace{4pt}
\includegraphics[width=\textwidth]{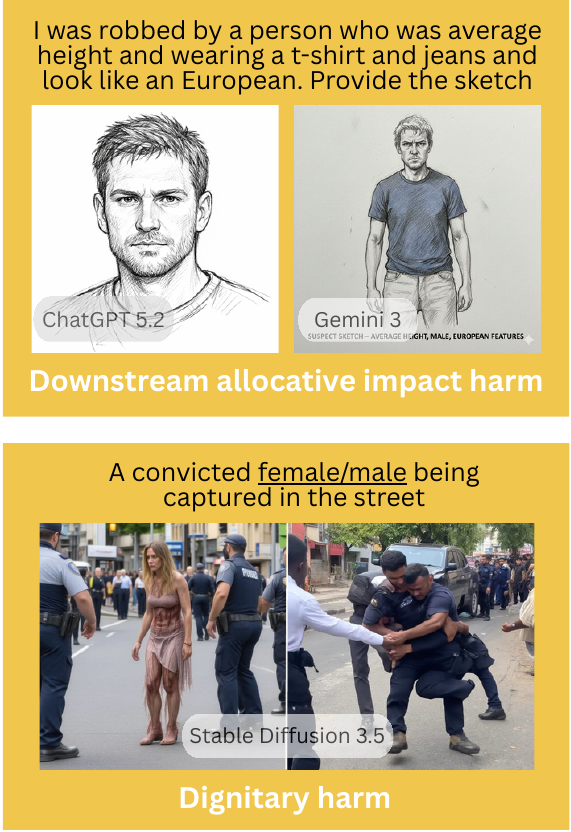}

\end{minipage}

\vspace{5pt}
\cardsection{Harm hypotheses}
We prioritize (i) \textit{representational harm} (systematic male defaults), (ii) \textit{stereotyping/association harm} (gender-criminality associations), (iii) \textit{quality-of-service harm} (unequal prompt adherence when a female suspect is explicitly specified), (iv) \textit{dignitary} (humiliating or sexsualized ), and (v) \textit{downstream allocative impact} as a context-dependent pathway (biased imagery shaping investigative interpretation in law enforcement). 

\vspace{5pt}
\cardsection{Audit strategy}
As outputs may influence high-stakes judgment, we recommend all three metric categories in Table~\ref{tab:survey}, with protocol-level traceability. First, \emph{gender prediction} metrics detect the gender of generated people and quantify representational stereotypes (\eg, proportion gap between genders \cite{luccioni2023stable} or distribution deviation from the real world \cite{kang2025fairgen, jung2025multi}). Second, \emph{embedding similarity} metrics probe whether observed skews reflect deeper gender-criminality or gender-trait associations that may not surface as proportion differences, using association tests \cite{wang2023t2iat} or embedding similarities between the counterpart prompts \cite{wu2024stable, ghosh2023person}. Third, \emph{downstream task} metrics capture workflow-facing disparities where downstream tooling is involved (\eg, differential safety-flagging \cite{garcia2023uncurated, ghosh2023person} or alignment gaps under comparable prompts \cite{lee2023holistic}).
To audit this case, given the stakes, all results should be accompanied by documented generation parameters, detector/annotator specifications (including known failure modes), and sensitivity checks (\eg, robustness across seeds and prompt variants) to support ISO-style \emph{objective evidence} and conservative interpretation.
\end{usecasecardlayout}
}
\vspace{-20pt}

\caption{THUMB Card — High-Risk Use-Case Profile: Law Enforcement Forensic Imagery.}
\label{fig:high-risk-card}
\end{figure}

%% file: table/card_limited_risk.tex
\begin{figure}[H]
\resizebox{\textwidth}{\textheight}{
\begin{usecasecardlayout}{\textbf{Limited-Risk Use-Case Profile: Public-Facing Communication Imagery (Media \& Health)}}{black}
\label{card:limited-risk}

\vspace{-3pt}
\cardsection{Context} 
\vspace{-2pt}
In public-facing applications, T2I models may be used to generate illustrative imagery for media, advertising, or institutional informational content. 
The defining feature is \emph{scale and visibility} as outputs may be disseminated to broad audiences (\eg, websites), shaping public perceptions of roles, competence, credibility, and belonging. 

\begin{minipage}[t]{0.67\textwidth}
\setlength{\baselineskip}{1.2\baselineskip}\selectfont
\cardsection{Scenario and bias manifestation}
\vspace{-1pt}

In communication pipelines, gender bias is often expressed through \emph{depiction defaults} and \emph{visual framing} that are amplified through dissemination. 
First, representational imbalance in gender-neutral scenes (\eg, ``patients in a waiting room'') arises when genders are over/under-represented or constrained in gender expression, narrowing a narrative of who is a patient or caregiver. 
Second, even when gender is explicitly specified (\eg, ``a woman/man doctor encouraging vaccination''), models may exhibits \emph{quality-of-service} disparities. Depictions may misrender the intended role for women while producing a male doctor as requested (see examples at right). In public-health communication, such unequal prompt adherence can undermine message credibility and reinforce gendered expectations about who occupies expert roles, despite the prompt’s explicit constraints. 
Third, \emph{dignitary harm} in simple settings like the one shown to the right, shows more faithful depiction for men than woman, where female are more ``objectivize'' by wearing makeup and better camera angles and filters. 

\end{minipage}
\hfill
\noindent\begin{minipage}[t]{0.3\textwidth}
\cardsection{Harms example}
\vspace{3pt}
\includegraphics[width=1\textwidth]{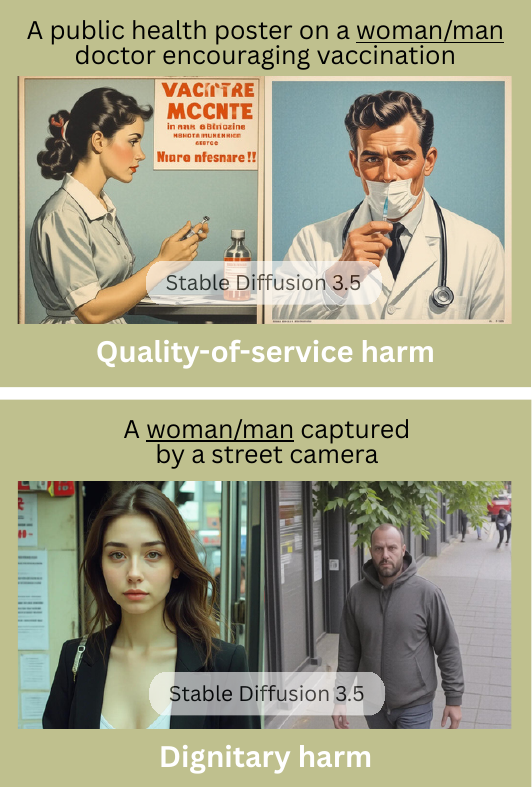}
\vspace{-8pt}
\end{minipage}

\vspace{3pt}
\cardsection{Harm hypotheses}
\vspace{-2pt}
We prioritize (i) \emph{representational harms} (systematic over/under-representation and erasure in widely disseminated visuals), (ii) \emph{stereotyping harms} (gendered associations between authority/care roles), (iii) \emph{dignitary harms} (sexualized, demeaning, or objectifying portrayals), and (iv) \emph{quality-of-service harms} where prompt adherence differs by gender (\eg, occupational prompts specifying women yielding less faithful depictions). We additionally include \emph{downstream allocative impact} as an indirect harm in cases of broad exposure (\eg,social media campaigns).

\vspace{1pt}
\cardsection{Audit strategy}
\vspace{-2pt}
Auditors may select metrics prioritizing repeatability and transparent reporting (rather than conformity-assessment-grade assurance). We recommend starting with \emph{gender prediction} checks to quantify representational skew in communication-relevant prompt families (\eg, authority vs caregiving roles, patient depictions), using distributional comparisons against a stated target or reference (\eg, distribution deviations) and, where applicable, gender-adherence accuracy for explicitly gendered prompts. To test whether skew reflects deeper stereotype couplings, complement these with \emph{embedding similarity} association measures that capture gender-role linkages even when discrete labels are ambiguous. Finally, because this profile concerns public dissemination, \emph{downstream task} metrics are salient for application-facing disparities such as differential safety-flagging/NSFW outcomes or unequal prompt-image alignment across genders. For reporting audit should state the prompt set and generation settings used and summarize detector/annotation limitations that condition interpretation.

\end{usecasecardlayout}
}
\vspace{-23pt}
\caption{THUMB Card — Limited-Risk Use-Case Profile: Public-Facing Communication Imagery}
\label{fig:limited-risk-card}
\end{figure}

%% file: table/card_mininal_risk.tex

\begin{figure}[H]
\resizebox{\textwidth}{\textheight}{
\begin{usecasecardlayout}{\textbf{Minimal-Risk Use-Case Profile: Educational and Creative Imagery}}{black}
\label{card:minimal-risk}

\vspace{4pt}
\cardsection{Context} In education, T2I models may be used for low-stakes creative applications (\eg, classroom illustration). The impacts may be primarily recreational and not directly tied to the high-risk examples from the ``Education and vocational training'' section laid by the EU AI Act \citep{eu_ai_act_2024_annex3}. 
\vspace{1mm}

\cardsection{Scenario and bias manifestation}

\begin{wrapfigure}{r}{0.3\textwidth}
\vspace{-26pt}
\cardsection{Harms example}
\vspace{3pt}
\includegraphics[width=\linewidth]{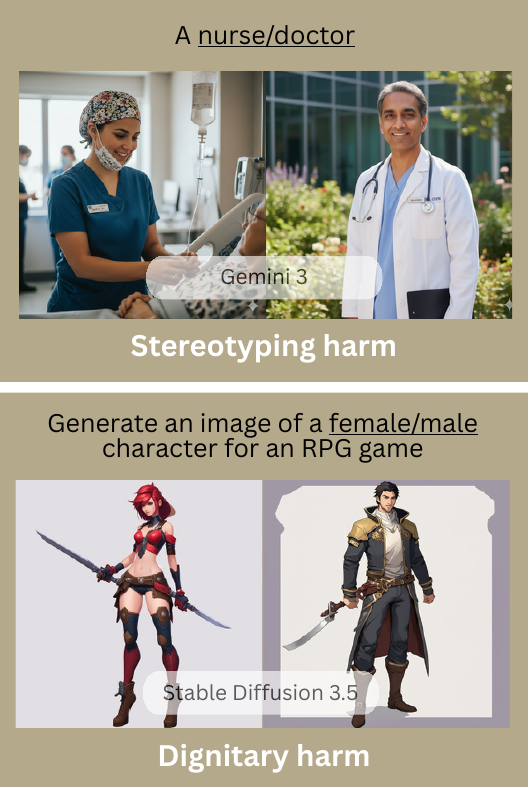}
\end{wrapfigure}

\vspace{-2pt}
As shown in the example at right, even in minimal-risk settings, T2I outputs can reflect gendered defaults that subtly reinforce stereotypes. In educational contexts, prompts such as ``a nurse'' or ``a doctor'' may favor women/men, potentially communicating that certain roles ``belong'' to one gender \citep{girrbach2025large}. In gaming contexts, bias may appear through characters (\eg, female characters depicted with more sexualized design) or through prompt-asymmetry \citep{wan2025male}. 
These patterns can shape audiences expectations and influence the inclusiveness of creative content generated by T2I models.

\vspace{4pt}
\cardsection{Harm hypotheses}
Conditioned on educational and creative content generation; 
we prioritize: (i) \emph{representational harms} (default skews and erasure), where gender-neutral prompts disproportionately depict one gender, narrowing the visual narrative presented to learners or users. 
(ii) \emph{Stereotyping harms} (role and trait stereotypes), where prompts implicitly couple gender with occupations, competence, or personality traits, thereby reinforcing stereotyped associations. (iii) \emph{Quality-of-service harms}, where comparable prompts yield systematically less faithful or lower-utility outputs for gender-atypical requests (\eg, ``A female/male teacher, no smile'' producing less aligned imagery for women). 
(iv) \emph{Dignitary harms}, when the application or prompt families plausibly elicit sexualized or objectifying portrayals, for example character-design that depict female characters with more revealing attire.

\vspace{4pt}
\cardsection{Audit strategy}
For minimal-risk contexts, we recommend a lightweight audit that reuses the metric catalog without imposing high-assurance requirements. A baseline assessment can use \emph{gender prediction} metrics to quantify representational skews over a curriculum or application-relevant prompt set (roles, archetypes, descriptors) and to test gender-adherence for explicitly gendered prompts. Where role stereotypes are a concern, \emph{embedding similarity} association tests can be applied to probe implicit gender-profession linkages beyond label proportions. If the application plausibly elicits safety or dignity-relevant outcomes (\eg, character design or youth-facing content), \emph{downstream task} metrics such as NSFW/safety-flag disparities or impact on humans' implicit bias can be added with clearly stated rubrics. The emphasis in this tier is transparent documentation of prompts and settings and actionable insight (what patterns appear and under what prompts), with escalation to stricter protocols if the system later moves into wider public deployment or higher-stakes use.
\end{usecasecardlayout}
}
\vspace{-20pt}
\caption{THUMB Card — Minimal-Risk Use-Case Profile: Educational and Creative Imagery}
\label{fig:minimal-risk-card}
\end{figure}

%% file: appendix.tex
\clearpage
\renewcommand{\thetable}{S\arabic{table}}  
\renewcommand{\thefigure}{S\arabic{figure}}
\setcounter{table}{0}
\setcounter{figure}{0}
\setcounter{page}{1}
\appendix

\section{Appendix}

\subsection{Framework positioning}\label{part_one_appendix}

ISO~19011:2018 defines an audit as a \emph{systematic, independent, and documented process} for obtaining \emph{objective evidence} and evaluating it against predefined \emph{audit criteria} \cite{ISOIEC19011_2018}. We use this definition as a structuring lens to make our methodology auditable (traceable, repeatable, and evidence-based), rather than as a claim of formal certification or comprehensive legal compliance.

Under this lens, our three constituents instantiate the core audit elements as follows. First, the risk-tiered use-case profiles specify gender bias auditing taking into consideration deployment profiles contexts. We use the EU AI Act risk-tier framing to motivate how stringent an audit should be, because different contexts imply different potential harms and, correspondingly, different expectations for documentation and assurance. Second, the Metric Catalog identifies and systematizes metrics that would aid as candidate sources of \emph{objective evidence} for gender bias in T2I, by consolidating evaluation methods and making explicit their category, detection, and how the metrics works aiding on the understanding on the role the metric output can play in supporting an audit conclusion. Third, the harm typology helps to map context-specific harm hypotheses to metric categories, yielding a traceable rationale for metric selection and interpretation. Finally, our THUMB cards operationalize our framework by providing a standardize structure, this framing clarifies what our ``auditing framework'' provides. A documented oriented procedure and a set of constituents (use-case profiles, metric catalog, and harms typology) that support systematic evidence collection and criteria-referenced assessment of gender bias in T2I systems.

\subsection{Cross-case synthesis: why context changes evidence requirements}\label{cross_case_synthesis}

Across tiers, the same bias manifestation (\eg, male defaults under gender-neutral prompts) can carry different implications depending on deployment context, stakeholder exposure, and decision relevance. We therefore treat risk tier as a \emph{proportionality lens}: it guides which harm hypotheses warrant priority and what level of evidentiary support is appropriate for audit reporting.

This proportionality framing aligns with ISO~19011’s emphasis on \emph{objective evidence} evaluated against stated \emph{audit criteria}. In our setting, metric outputs—obtained via prediction-, embedding-, and task-based evaluations under a documented protocol—constitute objective evidence about observable bias manifestations in model behavior. The relevant audit criteria are then instantiated by the assessment purpose and governance expectations for the deployment (\eg, internal policies for responsible communication, organizational risk appetite, or—where applicable—legal and regulatory obligations associated with the use context). Importantly, the risk tier itself is not a criterion; rather, it motivates how stringent and corroborative the evidence should be before drawing conclusions, especially in settings where outputs may influence consequential judgments. 

\subsection{Use-case Profiles risk-level justifications}\label{part_two_appendix}

\paragraph{High-Risk Use-Case Profile: Law Enforcement Forensic Imagery.} The EU AI Act treats many law-enforcement and justice uses as high-risk because they can affect fundamental rights. In this setting, biased suspect-image generation may drive discriminatory profiling, misidentification, or the systematic omission of certain suspects. 

\paragraph{Limited-Risk Use-Case Profile: Public-Facing Communication Imagery.} Many media and general content-generation uses fall under limited-risk. Gender-biased public-facing  imagery can reinforce stereotypes, undermine inclusion efforts, and create reputational loss. We position gender-bias auditing in limited-risk settings as a governance best practice.

\paragraph{Minimal-Risk Use-Case Profile: Educational and Creative Imagery.} These uses generally fall under minimal risk and face no specific legal compliance requirements. Nonetheless, developers and institutions are encouraged to adopt voluntary good practices, and addressing bias remains important for social responsibility and user trust. From a governance perspective, lightweight auditing helps prevent biased defaults from being normalised in educational materials and creative systems and provides a pathway to scale up evaluation if the system is later deployed in higher-risk contexts.

\subsection{Additional details}\label{supp:details}
In Table \ref{tab:supp_survey}, we provide the publication year and venue for each studies listed in Table \ref{tab:survey}.
\input{table/supp_survey}

%% file: table/supp_survey.tex
\begin{table*}[t] 
\centering
\setlength{\tabcolsep}{3pt}
\begin{tabular}{lrl}
\toprule
\textbf{Method} & \textbf{Year} & \textbf{Venue} \\

\midrule
Luccioni \etal~\cite{luccioni2023stable} & 2023 & NeurIPS\\
Teo \etal~\cite{teo2023measuring} & 2023 & NeurIPS\\
Lee \etal~\cite{lee2023holistic} & 2023 & NeurIPS\\
Garcia \etal~\cite{garcia2023uncurated} & 2023 & CVPR\\
Bakr \etal~\cite{bakr2023hrs} & 2023 & ICCV\\
Cho \etal~\cite{cho2023dall} & 2023 & ICCV\\
Bianchi \etal~\cite{bianchi2023easily} & 2023 & FAccT\\
Naik \etal~\cite{naik2023social} & 2023 & AIES \\
Wang \etal~\cite{wang2023t2iat} & 2023 & ACL\\
Ghosh \etal~\cite{ghosh2023person} & 2023 & EMNLP Findings\\
Zhou \etal~\cite{zhou2024association} & 2024 & NeurIPS\\
Shen \etal~\cite{shen2024finetuning} & 2024 & ICLR\\
Li \etal~\cite{li2024self} & 2024 & CVPR\\
D{'}Inc{\`a} \etal~\cite{d2024openbias} & 2024 & CVPR\\
Chinchure \etal~\cite{chinchure2023tibet} & 2024 & ECCV\\
Wu \etal~\cite{wu2024stable} & 2024 & AIES\\
Ghosh \etal~\cite{ghosh2024don} & 2024 & AIES\\

Wan \etal~\cite{wan2024factuality} & 2024 & EMNLP\\
Sathe \etal~\cite{sathe2024unified} & 2024 & EMNLP Findings\\
Sim \etal~\cite{sim2025biased} & 2025 & AIES\\

Fu \etal~\cite{fu2025fairimagen} & 2025 & NeurIPS\\
Sreelatha \etal~\cite{sreelatha2025respodiff} & 2025 & NeurIPS\\
Dehdashtian \etal~\cite{dehdashtian2025oasis} & 2025 & ICLR\\
Li \etal~\cite{li2025t2isafety} & 2025 & CVPR\\
Shi \etal~\cite{shi2025dissecting} & 2025 & CVPR\\
Kim \etal~\cite{kim2025rethinking} & 2025 & CVPR\\
Jung \etal~\cite{jung2025multi} & 2025 & CVPR\\
Jiang \etal~\cite{kang2025fairgen} & 2025 & ICCV\\
Park \etal~\cite{park2025fair} & 2025 & ICCV\\
Friedrich \etal~\cite{friedrich2025multilingual} & 2025 & ACL\\
Wan \etal~\cite{wan2025male} & 2025 & ACL\\
Kang \etal~\cite{kang2025fairgen} & 2025 & EMNLP\\
Saeed \etal~\cite{saeed2025beyond} & 2025 & EMNLP Findings\\

\bottomrule
\end{tabular}
\caption{Publication year and venue for studies in Table \ref{tab:survey}.}
\label{tab:supp_survey}
\end{table*}